\documentclass{iau}
 \usepackage{graphicx}

\title[IAUS291.~~Is magnetar a fact or fiction to us?] 
{Is magnetar a fact or fiction to us?} 

\author[H. Tong \& R. X. Xu]  
{H. Tong$^1$
 \and R. X. Xu$^2$}

\affiliation{$^1$Xinjiang Astronomical Observatory, Chinese Academy of Sciences,
Urumqi, Xinjiang 830011, China; Email: {\tt tonghao@xao.ac.cn} \\[\affilskip]
$^2$School of Physics and State Key Laboratory of Nuclear Physics
and Technology, Peking University, Beijing 100871, China; Email:
{\tt r.x.xu@pku.edu.cn}}

\pubyear{2012}
\volume{291}  
\jname{\mbox{Neutron Stars and Pulsars: Challenges and Opportunities after 80 years}}
\editors{J. van Leeuwen, ed.}
\begin{document}

\maketitle

\begin{abstract}

The key point of studying AXPs/SGRs (anomalous X-ray pulsars/soft
gamma-ray repeaters) is relevant to the energy budget. Historically,
rotation was thought to be the only free energy of pulsar until the
discovery of accretion power in X-ray binaries. AXPs/SGRs could be
magnetars if they are magnetism-powered, but would alternatively be
quark-star/fallback-disk systems if more and more observations would
hardly be understood in the magnetar scenario.

\end{abstract}


\firstsection 
\section{Introduction}

Anomalous X-ray pulsars/soft gamma-ray repeaters (AXPs/SGRs) are
focused because of their huge energy release and peculiar
behavior, suggesting that extra energy sources besides spin and
accretion powers should play an important role there.
Magnetic energy would be one of the candidates, which was initially
proposed. However, this viewpoint could be challenged by more and
more observations.
It is worth noting here that, to solve the energy budget would be a
key to understand the nature of compact stars, the equation state of
dense matter at supra-nuclear density.

Let's have a brief note on the history.
Rapid rotation was generally thought to be the only energy source
for pulsar emission soon after the discovery of radio pulsars
(\cite{mt77}) until the discovery of accretion-powered pulsars in
X-ray binaries (\cite{pr72}).
However, AXPs/SGRs have long spin periods (thus low spindown power,
their X-ray luminosities are much larger than their spindown powers)
and no binary companions, which rules out spin and accretion in
binary system as the power sources for the emission.
It was then proposed that SGR-like bursts as well as the persistent
X-ray emission could plausibly be the result of field decay of
ultra-magnetic neutron stars if MHD dynamo action in the proto-stars
is very effective in case that the objects spin initially at periods
of $\sim 1$\,ms (e.g., \cite{DT92}).
Because of the starquakes in the crusts of normal neutron stars, a
self-induction electric field is created. The strong electric field
could initiate avalanches of pair creation in the magnetosphere and
certainly accelerate particles, resulting in a so-called {\em
magnetar corona} (\cite{bt06}), from which high-energy bursts could
be observed. The power source of AXPs/SGRs is actually the magnetic
energy through field-reconnection there. This magnetar model is very
popular nowadays in the astrophysical community.

Unfortunately, many predictions (see following sections) in the
magnetar model have {\em never} been confirmed with certainty yet by
later observations, and we then have to revisit the real nature of
AXP/SGRs: are they really magnetars or alternatives?
We are repeatedly asking these questions (\cite{xu07}, \cite{T11}),
and try to conclude that AXP/SGRs might not be magnetars but could
be quark-star/fallback-disk systems.

\section{Evidence for magnetars?}

The periods and period derivatives of different kinds of pulsar-like
stars are shown in Fig.~1.
\begin{figure}[!htbp]
 \centering
\includegraphics[width=0.75\textwidth]{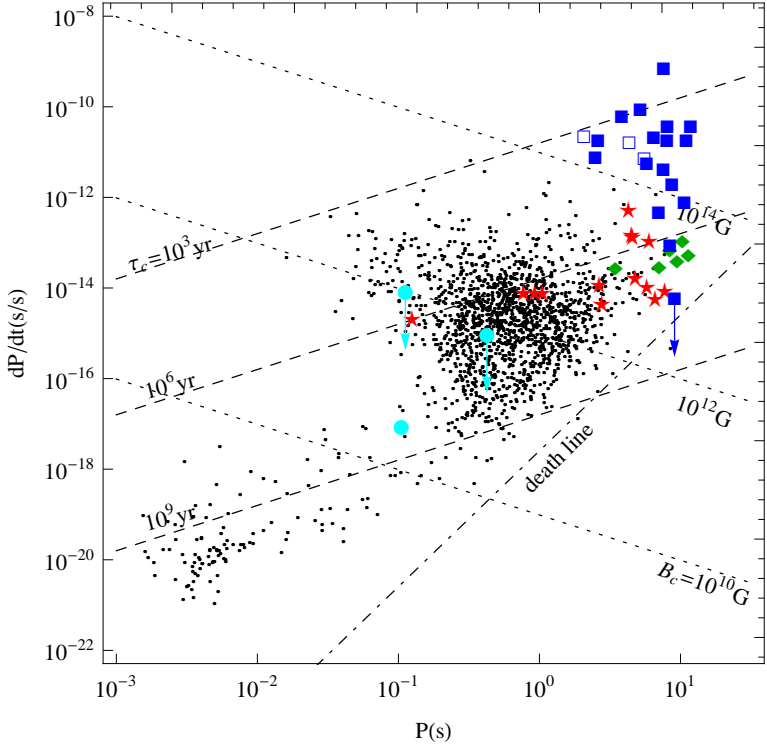}
\caption{P-$\mathrm{\dot{P}}$ diagram. Squares are for AXPs and
SGRs, empty squares for the radio loud magnetars (from
McGill online catalog), diamonds for X-ray dim isolated neutron stars
(XDINSs), stars for rotating radio transients (RRATs), 
filled circles for central compact objects (CCOs), and dots
for normal pulsars (from the ATNF).}%
\label{f1}
\end{figure}
We are summarizing possible observational evidence for magnetars
below.
\begin{itemize}
\item Strong dipole magnetic B-fields, measured by assuming magnetic dipole
braking.
\item Surface B-fields inferred from cyclotron absorptions if protons are
responsible.
\item Magnetic confinement of the giant flare tails after sudden release of magnetic
energy.
\item Magnetic suppression of scattering cress-section account for luminosity $L_x\sim 10^7 L_{\rm Edd}$.
\item SGR-like bursts from the high-B PSR J1846-0258 (but, how about the low-B
SGRs?).
\item Energy of both persistent \& burst emissions, spectral modeling (but
model-dependent).
\end{itemize}

\section{Challenges to the magnetar idea}

%
\begin{itemize}
\item Energetic supernova remnants associated with magnetars due to initial faster
spins (initial period $P_0\sim 1$\,ms) and higher B-fields ($B_0\sim 10^{14-16}$\,G).
\item A proto-neutron star with small $P_0$ and high $B_0$ may result in a large kick
velocity.
\item No radio emissions because of high B-field (but transient emissions are
detected).
\item Energetic gamma-rays from outer gaps, to be detectable by Fermi-LAT.
\item The unexpected discovery of a low-B SGRs, $B<7\times 10^{12}$\,G.
\item Unexpected existence of transient magnetars and high-B PSRs.
\item Magnetar free precession caused by higher mountains due to higher B-field.
\end{itemize}

\section{A solution?}

In view of the failed predictions of the magnetar model, listed in
\S3, we are obliged to think about alternative scenarios of
AXP/SGRs.
We note that the peculiar manifestations of AXP/SGRs would relate
closely their inner structures, i.e., the physics of dense matter at
supranuclear density.
It is well known that baryons of an evolved massive star should be
significantly compressed during core-collapse supernova, but the
nature of this compressed baryonic matter is still a matter of
debate due to strong non-perturbative effects of the fundamental
color interaction.
At a few nuclear densities, neutron stars containing almost only neutrons
(and other hadrons) are presumed to born soon after supernova, while
the remnant core might be composed by quark matter if the quark
degrees of freedom would not be negligible there.
Nonetheless, an emergence of quark-cluster state would be possible
if the coupling between quarks inside compact stars is still very
strong, forming a solid quark matter star (\cite{xu03}).
An accretion-induced quake model for AXP/SGRs is then proposed
(\cite{xty06}), in which the energy release during star quakes can
be estimated as high as %
$%
\Delta E =(G M^2/R)(\Delta R/R) \sim 5\times 10^{47}|\Delta
R/R|/(10^{-6})
$ ergs, %
to be enough to power the SGR giant flares.

Strong ejection (or wind) would certainly occur during
accretion-induced energy release and/or a star quake-induced burst,
and the central star may undergo a period of wind braking.
If AXP/SGRs are braked by wind instead of magnetic dipole radiation,
then their magnetosphere structure is different from that of normal
pulsars. This may explain the non-detection in {\it Fermi}
observations of magnetars (\cite{Tong2012}).
The extended emission around AXP 1E 1547.0-5408 may be a
magnetism-powered pulsar wind nebula. Under the wind braking
scenario, a braking index smaller than three is expected.

How can one finally differentiate the magnetar model and the
quark-star/fallback-disk model?
X-ray polarimetry may play an important role in identifying the real
equation of state of dense matter at supranuclear density
(\cite{lfx12}). We are expecting such an advanced facility to give a
final answer to this 80-year-longstanding problem.

\section{Conclusions}

Both the magnetar model and the fallback disk model for AXP/SGRs are
discussed. AXP/SGRs could be magnetars, but the origin and presence
of strong dipole field are challenged by recent observations.
Alternatively, both the bursts and persistent emissions of AXP/SGRs
could be understandable in the quark-star/fallback-disk model.

\vspace{2mm}%
\noindent %
{\bf Acknowledgments}.
This work is supported by 973-projects (2012CB821800, 2009CB82-
4800), NSFC (Grant Nos. 11103021, 10935001, 10973002), the Programme
of the Light in China's Western Region and the John Templeton
Foundation.


\end{document}